\documentclass{article}
\usepackage{amsmath}
\usepackage{graphicx}

\begin{document}

\textbf{The generalized Uhlenbeck-Goudsmit hypothesis:}

`\textbf{magnetic' }$S^{a}$ \textbf{and `electric' }$Z^{a}$ \textbf{spins}%
\bigskip \bigskip

Tomislav Ivezi\'{c}

Ru\mbox
{\it{d}\hspace{-.15em}\rule[1.25ex]{.2em}{.04ex}\hspace{-.05em}}er Bo\v{s}%
kovi\'{c} Institute, PO Box 180, 10002 Zagreb, Croatia

E-mail: ivezic@irb.hr\textit{\bigskip \bigskip }

\noindent \textbf{Abstract}

\noindent In this paper, the connection between the dipole moment tensor $%
D^{ab}$ and the spin four-tensor $S^{ab}$ is formulated in the form of the
generalized Uhlenbeck-Goudsmit hypothesis, $D^{ab}=g_{S}S^{ab}$. It is also
found that the spin four-tensor $S^{ab}$ can be decomposed into two
4-vectors, the usual `space-space'\ intrinsic angular momentum $S^{a}$,
which will be called `magnetic' spin (mspin), and a new one, the
`time-space'\ intrinsic angular momentum $Z^{a}$, which will be called
`electric' spin (espin). Both spins are equally good physical quantities.
Taking into account the generalized Uhlenbeck-Goudsmit hypothesis, the
decomposition of $S^{ab}$ and the decomposition of $D^{ab}$ into the dipole
moments $m^{a}$ and $d^{a}$, we find that an electric dipole moment (EDM) of
a fundamental particle, as a four-dimensional (4D) geometric quantity, is
determined by $Z^{a}$ and not, as generally accepted, by the spin $\mathbf{S}
$ as a 3-vector. Also it is shown that neither the $T$ inversion nor the $P$
inversion are good symmetries in the 4D spacetime. In this geometric
approach, only the world parity $W$, $Wx^{a}=-x^{a}$, is well defined in the
4D spacetime. Some consequences for elementary particle theories and
experiments that search for EDM are briefly discussed. \bigskip

\noindent \textit{Henceforth space by itself, and time by itself, are doomed
\newline
to fade away into mere shadows and only a kind of union of \newline
the two will preserve an independent reality.}

\noindent H. Minkowski\bigskip \bigskip

\noindent PACS numbers: 03.30.+p, 13.40.Em, 11.30.Er, 03.65.Sq \bigskip
\bigskip \medskip

\noindent \textbf{1. Introduction}\bigskip

\noindent In this geometric approach, it is considered that an independent
physical reality, as in Minkowski's statement above, is attributed to the
geometric quantities that are defined on the four-dimensional (4D) spacetime
and not, as usually accepted, by the 3-vectors. Such geometric quantities
are introduced in section 2. There, using a general rule for the
decomposition of a second rank antisymmetric tensor, the dipole moment
tensor $D^{ab}$ is decomposed into the electric dipole moment (EDM) $d^{a}$
and the magnetic dipole moment (MDM) $m^{a}$ (\ref{d}). The main results are
obtained in section 3. Using the same rule, it is shown that the spin
four-tensor $S^{ab}$ can be decomposed into two 4-vectors, the usual
`space-space'\ intrinsic angular momentum $S^{a}$, which will be called
`magnetic' spin (mspin), and a new one, the `time-space'\ intrinsic angular
momentum $Z^{a}$, (\ref{es}), which will be called `electric' spin (espin).
Then, the connection between $D^{ab}$ and $S^{ab}$ is formulated in the form
of the generalized Uhlenbeck-Goudsmit hypothesis, $D^{ab}=g_{S}S^{ab}$ (\ref%
{gu}). It is shown in (\ref{dm}) that an EDM of a fundamental particle, such
as a 4-vector, is determined by the espin $Z^{a}$ and not by the spin $%
\mathbf{S}$. (The usual 3-vectors will be designated in bold-face.) The
relation (\ref{dm}) also shows that the MDM of a fundamental particle is
determined by the mspin $S^{a}$. In section 4, it is proved that neither the
$T$ inversion nor the $P$ inversion are good symmetries in the 4D spacetime.
In this geometric approach, only the world parity $W$, $Wx^{a}=-x^{a}$, is
well defined in the 4D spacetime. Hence, in this approach, the existence of
an EDM is not connected in any way with $T$ violation or, under the
assumption of CPT invariance, with CP violation. In section 5, the results
obtained are used to discuss recent experimental searches for a permanent
EDM of particles, and different shortcomings in the interpretations of the
results of measurements are considered. The results obtained in this paper
significantly differ from the usual formulation. However, this is a
consistent theory with intrinsically covariant objects. It could be
considered as an `alternative' but viable (in my opinion) view of the
intrinsic angular momentums-spins and the associated dipole moments of the
elementary particles. The treatment of the Trouton - Noble experiment [1]
with the angular momentum four-tensor $M^{ab}$ and the torque four-tensor $%
N^{ab}$ provides important supporting evidence for the formulation presented
here. More supporting evidence comes from the resolution of Jackson's
paradox, which is obtained in [2] using the same geometric quantities $%
M^{ab} $ and $N^{ab}$ as those in [1]. \bigskip \bigskip

\noindent \textbf{2. The 4D geometric approach}\bigskip

\noindent We shall deal with 4D geometric quantities that are defined
without reference frames, e.g. the 4-vectors of the electric and magnetic
fields $E^{a}$ and $B^{a}$, the electromagnetic field tensor $F^{ab}$, the
dipole moment tensor $D^{ab}$, the 4-vectors of the EDM $d^{a}$ and the MDM $%
m^{a}$, etc. In the following, we shall rely on the results and the
explanations from [3]; see also references therein. As stated in [3],
according to [4], $F^{ab}$ can be taken as the primary quantity for the
whole of electromagnetism. $E^{a}$ and $B^{a}$ are then derived from $F^{ab}$
and the 4-velocity of the observers $v^{a}$:%
\begin{eqnarray}
F^{ab} &=&(1/c)(E^{a}v^{b}-E^{b}v^{a})+\varepsilon ^{abcd}v_{c}B_{d},  \notag
\\
E^{a} &=&(1/c)F^{ab}v_{b},\ B^{a}=(1/2c^{2})\varepsilon ^{abcd}F_{bc}v_{d};\
E^{a}v_{a}=B^{a}v_{a}=0.  \label{f}
\end{eqnarray}%
The frame of `fiducial'\ observers is the frame in which the observers who
measure $E^{a}$ and $B^{a}$ are at rest. That frame with the standard basis $%
\{e_{\mu }\}$ in it is called the $e_{0}$-frame. (The standard basis $%
\{e_{\mu };\ 0,1,2,3\}$ consists of orthonormal 4-vectors with $e_{0}$ in
the forward light cone. It corresponds to the specific system of coordinates
with Einstein's synchronization [5] of distant clocks and Cartesian space
coordinates $x^{i}$.) In the $e_{0}$-frame, $v^{a}=ce_{0}$, which, with (\ref%
{f}), yields $E^{0}=B^{0}=0$ and $E^{i}=F^{i0}$, $B^{i}=(1/2c)\varepsilon
^{ijk0}F_{jk}$. Therefore $E^{a}$ and $B^{a}$ can be called the
`time-space'\ part and the `space-space'\ part, respectively, of $F^{ab}$.
The reason for the quotation marks in `time-space'\ \ and `space-space'\
will be explained in section 4.

As proved in, e.g., [6], any second rank antisymmetric tensor can be
decomposed into two 4-vectors and a unit time-like 4-vector (the
4-velocity/c). This rule can be applied to $D^{ab}$. As shown in [3], $%
D^{ab} $ is the primary quantity for dipole moments. Then $d^{a}$ and $m^{a}$
are derived from $D^{ab}$ and the 4-velocity of the particle $u^{a}$:%
\begin{eqnarray}
D^{ab} &=&(1/c)(d^{a}u^{b}-d^{b}u^{a})+(1/c^{2})\varepsilon
^{abcd}m_{c}u_{d},  \notag \\
m^{a} &=&(1/2)\varepsilon ^{abcd}D_{cb}u_{d},\ d^{a}=(1/c)D^{ab}u_{b},
\label{d}
\end{eqnarray}%
with $d^{a}u_{a}=m^{a}u_{a}=0$. In the particle's rest frame (the $K^{\prime
}$ frame) and the $\{e_{\mu }^{\prime }\}$ basis, $u^{a}=ce_{0}^{\prime }$,
which, with (\ref{d}), yields that $d^{\prime 0}=m^{\prime 0}=0$, $d^{\prime
i}=D^{\prime i0}$, $m^{\prime i}=(c/2)\varepsilon ^{0ijk}D_{jk}^{\prime }$.
Therefore $d^{a}$ and $m^{a}$ can be called the `time-space'\ part and the
`space-space'\ part, respectively, of $D^{ab}$.

In this geometric approach, the interaction term in the Lagranian for the
interaction between $F^{ab}$ and $D^{ab}$ can be written as a sum of two
terms [3]:
\begin{eqnarray}
(1/2)F_{ab}D^{ba}
&=&(1/c^{2})[-(E_{a}d^{a}+B_{a}m^{a})(v_{b}u^{b})+(E_{a}u^{a})(v_{b}d^{b})
\label{i} \\
&&+(B_{a}u^{a})(v_{b}m^{b})]-(1/c^{3})[\varepsilon
^{abcd}(E_{b}m_{d}-c^{2}B_{b}d_{d})v_{a}u_{c}].  \notag
\end{eqnarray}%
Observe that every term on the rhs of (\ref{i}) contains both velocities $%
u^{a}$ and $v^{a}$. This fact differs (\ref{i}) from all previous
expressions for the interaction between dipole moments and the electric and
magnetic fields. As seen from the last two terms they naturally contain the
interaction of $E^{a}$ with $m^{a}$, and $B^{a}$ with $d^{a}$, which are
required for the explanations of the Aharonov-Casher effect and the R\"{o}%
ntgen phase shift [3,7], and also of different methods of measuring EDMs,
e.g. such methods as in [8]. Moreover, there is no need for any
transformation. We only need to choose the laboratory frame as our $e_{0}$%
-frame and then to represent $E^{a}$, $m^{a}$ and $B^{a}$, $d^{a}$ in that
frame.

Furthermore, it is shown in [2] and [1] that the angular momentum
four-tensor $M^{ab}$, given as $M^{ab}=x^{a}p^{b}-x^{b}p^{a}$ (i.e. in [2]
and [1], the bivector $M=x\wedge p$), can be decomposed into the
`space-space'\ angular momentum of the particle $M_{s}^{a}$ and the
`time-space'\ angular momentum $M_{t}^{a}$ (both with respect to the
observer with velocity $v^{a}$):%
\begin{eqnarray}
M^{ab} &=&(1/c)[(M_{t}^{a}v^{b}-M_{t}^{b}v^{a})+\varepsilon
^{abcd}M_{s,c}v_{d}],  \notag \\
M_{s}^{a} &=&(1/2c)\varepsilon ^{abcd}M_{cb}v_{d},\
M_{t}^{a}=(1/c)M^{ab}v_{b},  \label{em}
\end{eqnarray}%
with $M_{s}^{a}v_{a}=M_{t}^{a}v_{a}=0$. $M_{s}^{a}$ and $M_{t}^{a}$ depend
not only on $M^{ab}$ but also on $v^{a}$. Only in the $e_{0}$-frame $%
M_{s}^{0}=M_{t}^{0}=0$ and $M_{s}^{i}=(1/2)\varepsilon ^{0ijk}M_{jk}$, $%
M_{t}^{i}=M^{i0}$. $M_{s}^{i}$ and $M_{t}^{i}$ correspond to the components
of $\mathbf{L}$ and $\mathbf{K}$ that are introduced, e.g., in [9]. However
Jackson [9], as all others, considers that only $\mathbf{L}$ is a physical
quantity whose components transform according to equation (11) in [9], which
we write as
\begin{equation}
L_{x}^{\prime }=L_{x},\ L_{y}^{\prime }=\gamma (L_{y}+\beta K_{z}),\
L_{z}^{\prime }=\gamma (L_{z}-\beta K_{y})  \label{el}
\end{equation}%
(for the boost in the $+x$ - direction). The components of $\mathbf{B}$ (and
of $\mathbf{E}$) are transformed in the same way (see equation (11.148) in
[10]), for example,
\begin{equation}
B_{x}^{\prime }=B_{x},\ B_{y}^{\prime }=\gamma (B_{y}+\beta E_{z}),\
B_{z}^{\prime }=\gamma (B_{z}-\beta E_{y}).  \label{bx}
\end{equation}%
\emph{The essential point is that in both equations,} (\ref{el}) \emph{and} (%
\ref{bx}), \emph{the transformed components,} $L_{i}^{\prime }$\emph{\ and }$%
B_{i}^{\prime }$\emph{,} \emph{are expressed by the mixture of components,} $%
L_{k}$\emph{,} $K_{k}$ \emph{and }$B_{k}$\emph{,} $E_{k},$\emph{\
respectively.}

Recently [11], a fundamental result was achieved that the usual
transformations (UT) of $\mathbf{E(r},t\mathbf{)}$, $\mathbf{B(r},t\mathbf{)}
$, equations (11.148) and (11.149) in [10], differ from the Lorentz
transformations (LT) (boosts) of the 4D geometric quantities that represent
the electric and magnetic fields.

Also, it is worth mentioning an important result regarding the usual
formulation of electromagnetism (as in [10]), which is presented in [12] and
discussed in [13]. It is explained in [12] that the usual $\mathbf{E(r,}t%
\mathbf{)}$, $\mathbf{B(r,}t\mathbf{)}$ are not correctly defined as the
quantities which have, in some basis of the 3D space, only three components,
since they are space- and time-dependent quantities. This means that they
are defined on the spacetime and that fact determines that such vector
fields, when represented in some basis, have to have four components (some
of them can be zero). It is argued in [12] that an individual vector has no
dimension; the dimension is associated with the vector space and with the
manifold where this vector is tangent. Hence, what is essential for the
number of components of a vector field is the number of variables on which
that vector field depends, i.e., the dimension of its domain. Thus, strictly
speaking, the time-dependent $\mathbf{E(r,}t\mathbf{)}$ and $\mathbf{B(r,}t%
\mathbf{)}$ cannot be the 3-vectors, since they are defined on the
spacetime. Therefore, from now on, we shall use the term `vector' for a
geometric quantity, which is defined on the spacetime and which always has
in some basis of that spacetime, e.g. the standard basis $\{e_{\mu }\}$,
four components (some of them can be zero). This refers to $m^{a}$, $d^{a}$,
$M_{s}^{a}$, $M_{t}^{a}$, ... as well. (In the preceding text, they are
called the 4-vectors.) However, an incorrect expression, the 3-vector, will
still remain for the usual $\mathbf{E(r,}t\mathbf{)}$, $\mathbf{B(r,}t%
\mathbf{)}$, $\mathbf{L}$, $\mathbf{K}$, the spin $\mathbf{S}$, etc.

For the `fiducial'\ observers, $v^{\mu }=ce_{0}^{\mu }$ and $E^{\mu }=F^{\mu
\nu }e_{0,\nu }$. As shown in [11], and also in [13], both the field $F^{ab}$
and the velocity (/c) of the `fiducial'\ observer have to be transformed by
the LT. This correct mathematical procedure yields that the components (in
the standard basis) $E^{\mu }$ transform by the LT as

\begin{equation}
E^{\prime 0}=\gamma (E^{0}-\beta E^{1}),\ E^{\prime 1}=\gamma (E^{1}-\beta
E^{0}),\ E^{\prime 2,3}=E^{2,3},  \label{be}
\end{equation}%
for the boost in the $+x^{1}$ - direction. Of course, this is the way in
which the components (in the standard basis) of any vector transform under
the LT. Hence, the same transformations as (\ref{be}) hold for the
components $B^{\mu }$, $M_{s}^{\mu }$, $d^{\mu }$, $m^{\mu }$, $S^{\mu }$, $%
Z^{\mu }$, etc. As noted in [12], and discussed in [13], Minkowski, in
section 11.6 in [14], was the first who correctly transformed the electric
and magnetic vectors.

\emph{The fundamental difference between the correct LT} \emph{(\ref{be}) of
the components (in the standard basis) and the UT (\ref{el}), (\ref{bx}) is
that the components} $E^{\mu }$, i.e. $B^{\mu }$, $M_{s}^{\mu }$, ..., \emph{%
transform by the LT again to the components} $E^{\prime \mu }$, i.e. $%
B^{\prime \mu }$,\emph{\ }$M_{s}^{\prime \mu }$, \emph{\ ..., respectively;
there is no mixing of components. }

As said in section 1, it is proved in [1] that the treatment with $M^{ab}$
(or $M_{s}^{a}$ and $M_{t}^{a}$) and the torque four-tensor $N^{ab}$ (or
vectors $N_{s}^{a}$ and $N_{t}^{a}$) is in true agreement (independent of
the chosen inertial reference frame and of the chosen system of coordinates
in it) with the Trouton-Noble experiment. Similarly, in [2] it is shown that
in such an approach with $M^{ab}$ and $N^{ab}$ the principle of relativity
is naturally satisfied and there is no Jackson's paradox. The true agreement
with experiments, when using 4D geometric quantities, is also obtained in
the second paper in [11] (the motional electromotive force), in the third
paper in [11] (the Faraday disc) and also in [15] (the well-known
experiments: the `muon'\ experiment, the Michelson-Morley - type
experiments, the Kennedy-Thorndike - type experiments and the Ives-Stilwell
- type experiments). This true agreement with experiments directly proves
the physical reality of the 4D geometric quantities. It is also shown in the
mentioned papers ([1], [11] and [15]) that the agreement between the
experiments that test special relativity and Einstein's formulation of
special relativity [5], which deals with the synchronously defined spatial
length , i.e. the Lorentz contraction, with the conventional dilatation of
time and also with the UT of the components of the 3-vectors $\mathbf{E}$
and $\mathbf{B}$, is not a true agreement since it depends on the chosen
synchronization, e.g. Einstein's synchronization or a drastically different,
nonstandard, radio (`r') synchronization; see also [16] and section 4 here.

\bigskip \bigskip

\noindent \textbf{3. The generalized Uhlenbeck-Goudsmit hypothesis;}

\textbf{`time-space'\ intrinsic angular momentum and the intrinsic
EDM\bigskip }

\noindent The above consideration can be directly applied to the \emph{%
intrinsic} angular momentum, the spin of an elementary particle. In the
usual approaches, e.g. section 11.11 A in [10], the relativistic
generalization of the spin $\mathbf{S}$ from a 3-vector in the particle's
rest frame is obtained in the following way: `The spin 4-vector $S^{\alpha }$
is the dual of the tensor $S^{\alpha \beta }$ in the sense that $S^{\alpha
}=(1/2c)\varepsilon ^{\alpha \beta \gamma \delta }u_{\beta }S_{\gamma \delta
}$, where $u^{\alpha }$ is the particle's 4-velocity.'\ The whole discussion
above about $F^{ab}$, $D^{ab}$ (\ref{d}) and particularly about $M^{ab}$ (%
\ref{em}) (spin is also an angular momentum) implies a more general
geometric formulation of the spin of an elementary particle. In analogy with
[1] and [2], we conclude that the primary quantity \emph{with} \emph{%
definite physical reality} for the \emph{intrinsic} angular momenta is the
spin four-tensor $S^{ab}$, which can be decomposed into two vectors, namely
the usual `space-space'\ intrinsic angular momentum $S^{a}$ and the
`time-space'\ intrinsic angular momentum $Z^{a}$:
\begin{eqnarray}
S^{ab} &=&(1/c)[(Z^{a}u^{b}-Z^{b}u^{a})+\varepsilon ^{abcd}S_{c}u_{d}],
\notag \\
S^{a} &=&(1/2c)\varepsilon ^{abcd}S_{cb}u_{d},\ Z^{a}=(1/c)S^{ab}u_{b},
\label{es}
\end{eqnarray}%
where $u^{a}=dx^{a}/d\tau $ is the velocity of the particle and it holds
that $S^{a}u_{a}=Z^{a}u_{a}=0$. $S^{a}$ and $Z^{a}$ depend not only on $%
S^{ab}$ but on $u^{a}$ as well. Only in the particle's rest frame, the $%
K^{\prime }$ frame, and the $\{e_{\mu }^{\prime }\}$ basis, $%
u^{a}=ce_{0}^{\prime }$ and $S^{\prime 0}=Z^{\prime 0}=0$, $S^{\prime
i}=(1/2c)\varepsilon ^{0ijk}S_{jk}^{\prime }$, $Z^{\prime i}=S^{\prime i0}$.
The definition (\ref{es}) essentially changes the usual understanding of the
spin of an elementary particle. It introduces a new `time-space'\ spin $%
Z^{a} $, which is a physical quantity in the same measure as it is the usual
`space-space'\ spin $S^{a}$.

In [17] it is asserted: `For an elementary particle, the only intrinsic
direction is provided by the spin $\mathbf{S}$. Then its intrinsic $\mathbf{%
\mu }=\gamma _{S}\mathbf{S}$ and its intrinsic $\mathbf{d}=\delta _{S}%
\mathbf{S}$, where $\delta _{S}$ is a constant.'\ (In [17] the unprimed
quantities are in the particle's rest frame.) Thus, both the MDM $\mathbf{m}%
^{\prime }$ and the EDM $\mathbf{d}^{\prime }$ (our notation) of an
elementary particle are determined by the usual spin $\mathbf{S}^{\prime }$.
In the usual approaches such a result is expected because only the
`space-space'\ intrinsic angular momentum is considered to be a well-defined
physical quantity. In contrast to [17] and other usual approaches, we
consider that the intrinsic direction in the 3D space is not important in
the 4D spacetime, since it does not correctly transform under the LT. As
already explained, in this geometric approach a definite physical reality is
attributed to $S^{ab}$ or to $S^{a}$ and $Z^{a}$ taken together (see (\ref%
{es})) in the same way as holds for the angular momentum four-tensor $M^{ab}$
and the angular momenta $M_{s}^{a}$ and $M_{t}^{a}$ (\ref{em}) (see [1] and
[2]).

Furthermore, in the usual approaches, there is a connection between the
magnetic moment $\mathbf{m}$ and the spin $\mathbf{S}$, $\mathbf{m}=\gamma
_{S}\mathbf{S}$. This is the well-known Uhlenbeck-Goudsmit hypothesis [18].
The whole of the above consideration suggests that instead of the above
connection between the 3-vectors $\mathbf{m}$ and $\mathbf{S}$ we need to
have the connection between the dipole moment tensor $D^{ab}$ and the spin
four-tensor $S^{ab}$. Obviously, it has to be formulated in the form of the
generalized Uhlenbeck-Goudsmit hypothesis as
\begin{equation}
D^{ab}=g_{S}S^{ab}.  \label{gu}
\end{equation}%
Taking into account the decompositions of $D^{ab}$ (\ref{d}) and $S^{ab}$ (%
\ref{es}), we find the connections between the dipole moments $m^{a}$ and $%
d^{a}$ and the corresponding intrinsic angular momenta $S^{a}$ and $Z^{a}$,
respectively, in a form that essentially differs from all usual approaches,
e.g. [17]:
\begin{equation}
m^{a}=cg_{S}S^{a},\ d^{a}=g_{S}Z^{a}.  \label{dm}
\end{equation}%
In the particle's rest frame and the $\{e_{\mu }^{\prime }\}$ basis, $%
u^{a}=ce_{0}^{\prime }$ and $d^{\prime 0}=m^{\prime 0}=0$, $d^{\prime
i}=g_{S}Z^{\prime i}$, $m^{\prime i}=cg_{S}S^{\prime i}$. Comparing this
last relation with $\mathbf{m}=\gamma _{S}\mathbf{S}$, we see that $%
g_{S}=\gamma _{S}/c.$ Thus, the intrinsic MDM $m^{a}$ of an elementary
particle is determined by the mspin $S^{a}$, whereas the intrinsic EDM $%
d^{a} $ is determined by the espin $Z^{a}$; the names mspin and espin come
from the connections given by (\ref{dm}). The relations (\ref{gu}) and (\ref%
{dm}) say that any fundamental particle has not only the intrinsic MDM $%
m^{a} $, but also the intrinsic EDM $d^{a}$ whose magnitude is $(1/c)$ of
that for $m^{a}$. We repeat once again that, in this theory, the existence
of the intrinsic EDM $d^{a}$ is obtained from the assumption that the
primary quantities (with independent physical reality) are the spin
four-tensor $S^{ab}$ and the dipole moment tensor $D^{ab}$, which can be
decomposed according to relations (\ref{es}) and (\ref{d}), respectively.
Then the usual connection between the 3-vectors $\mathbf{m}$ and $\mathbf{S}$
is generalized to the relations (\ref{gu}) and (\ref{dm}). The EDM obtained
in this way is of quite different physical nature than in the elementary
particle theories, e.g. in the standard model and in supersymmetric (SUSY)
theories. There, an EDM is obtained by a dynamic calculation and it stems
from an asymmetry in the charge distribution inside a fundamental particle,
which is thought of as a charged cloud. Here, as already stated, the EDM $%
d^{a}$ (see (\ref{dm})) emerges from the connection with the intrinsic
angular momentum $Z^{a}$, i.e. from (\ref{gu}) and (\ref{d}), (\ref{es}).

Recently, I have become aware of some papers in which formally similar, but
really different, results have been obtained. Westpfahl's formulae (3.15)
and (3.15a) in [19], and equation (21) in [20] are, at first glance, very
similar to our equations (\ref{es}) and (\ref{gu}), respectively. The first
important difference is that our approach deals with 4D geometric quantities
that are defined without reference frames. Hence our equations hold for any
reference frame and for any chosen system of coordinates in it. Westpfahl
deals only with components implicitly taken in the standard basis; thus only
Einstein's synchronization is considered (see the next section for a
nonstandard synchronization). Furthermore, it can be easily seen from
Westpfahl's formulae (3.15d) and (3.15e) in [19] and his equations
(5)-(6\textquotedblright ) in [20] that the definitions of the quantities
entering into his equations (3.15) and (3.15a) in [19] and equation (21) in
[20] are very different from the quantities that can be obtained from the 4D
geometric quantities entering into (\ref{es}) and (\ref{gu}). However, for
comparison, one has to take only components in the standard basis from the
4D geometric quantities. In [21], an EDM of the electron is associated with
the infinitesimal generator of the Lorentz boost and, as usual, the MDM of
the electron is associated with the infinitesimal generator of the 3D
spatial rotation. This corresponds to our relation (\ref{dm}), but in [21]
the authors deal only with components implicitly taken in the standard basis
and finally with the 3-vectors, $\mathbf{E}$, $\mathbf{B}$, $\mathbf{p}$, $%
\mathbf{A}$, etc. Such an approach cannot work when some nonstandard
synchronization, e.g. the `r'\ synchronization from the next section, is
used. An interesting application to spintronics of that EDM of the electron
is presented in [21]. In our approach, the spin-orbit interaction is a part
of (\ref{i}) when (\ref{gu}) and (\ref{dm}) are inserted in it. This will be
treated elsewhere.

In addition, let us find the equation of motion for the spin four-tensor $%
S^{ab}$. The rest frame equation of motion for the usual spin $\mathbf{S}$
is given, e.g., by (11.101) or (11.155) in [10], which we write as
\begin{equation}
d\mathbf{S}^{\prime }/dt^{\prime }=\mathbf{m}^{\prime }\times \mathbf{B}%
^{\prime }=\gamma _{S}\mathbf{S}^{\prime }\times \mathbf{B}^{\prime },
\label{bs}
\end{equation}%
where all the quantities are in the particle's rest frame, the $K^{\prime }$
frame, and the Uhlenbeck-Goudsmit hypothesis [18], $\mathbf{m}=\gamma _{S}%
\mathbf{S}$, is used. In Sec. 11.11 A in [10], a covariant generalization of
(\ref{bs}) is presented and it refers to the spin 4-vector $S^{\mu }$
(components in the $\{e_{\mu }\}$ basis). However, in the geometric approach
presented here, the relation $\mathbf{m}=\gamma _{S}\mathbf{S}$ is
generalized, equation (\ref{gu}), replacing the 3-vectors $\mathbf{m}$ and $%
\mathbf{S}$ by the dipole moment tensor $D^{ab}$ and the spin four-tensor $%
S^{ab}$, respectively. In the same way we can generalize (\ref{bs}). Using (%
\ref{gu}), the generalized equation of motion for the spin four-tensor $%
S^{ab}$ becomes
\begin{equation}
dS^{ab}/d\tau
=F^{ac}g_{cd}D^{db}-F^{bc}g_{cd}D^{da}=g_{S}[F^{ac}g_{cd}S^{db}-F^{bc}g_{cd}S^{da}].
\label{sg}
\end{equation}%
where $g_{ab}$ is the metric tensor. Equation (\ref{sg}) is written with
primary 4D geometric quantities, for the electromagnetic field $F^{ab}$ and
for dipole moments $D^{ab}$, i.e. for spins $S^{ab}$. (In [22], a very
similar equation of motion for the spin four-tensor $S^{\mu \nu }$ is
derived (see equation (14) there); however, Peletminskii and Peletminskii
[22] deal exclusively with components in the $\{e_{\mu }\}$ basis.) Of
course one can use the decompositions (\ref{f}) and (\ref{d}), or (\ref{es}%
), to obtain the generalized equation of motion (\ref{sg}) expressed in
terms of the fields $E^{a}$, $B^{a}$ and the dipole moments $d^{a}$, $m^{a}$
or the espin $Z^{a}$ and the mspin $S^{a}$. The consequences of (\ref{sg})
will not be investigated here, e.g. the generalization of the BMT equation
((11.164) in [10]), etc.. We remark only that (\ref{sg}) reduces to the
equation
\begin{equation}
dS^{\prime i}/d\tau =\varepsilon ^{0ijk}m_{j}^{\prime }B_{k}^{\prime
}=\gamma _{S}\varepsilon ^{0ijk}S_{j}^{\prime }B_{k}^{\prime }  \label{sm}
\end{equation}%
($cg_{S}=\gamma _{S}$) in the $K^{\prime }$ frame and the $\{e_{\mu
}^{\prime }\}$ basis. The $K^{\prime }$ frame is also chosen to be the $%
e_{0} $-frame, i.e. the observers who measure fields $E^{a}$ and $B^{a}$
move together with the dipole, $v^{a}=u^{a}=ce_{0}^{\prime }$ and
consequently $E^{\prime 0}=B^{\prime 0}=d^{\prime 0}=m^{\prime 0}=0$.
Furthermore, it is taken that in the $K^{\prime }$ frame $d^{\prime i}=0$,
i.e. $Z^{\prime i0}=0 $ and that $E^{\prime i}=F^{\prime i0}=0$. The
relation (\ref{sm}) corresponds to the equation with the 3-vectors (\ref{bs}%
). However, (\ref{sm}) is correctly expressed by the components of the 4D
geometric quantities, whereas (\ref{bs}) is written with the 3-vectors whose
transformations are not the LT and also it contains the coordinate time and
not the proper time.

Relations (\ref{es})-(\ref{dm}) with 4D geometric quantities $S^{ab}$, $%
S^{a} $ and $Z^{a}$, $D^{ab}$, $m^{a}$ and $d^{a}$ are fundamentally new
results that have not been mentioned previously in such a form in the
literature.

In addition, it is worthwhile to mention the classical references, [23], on
the relativistic theory of spin in classical electrodynamics. However, both
Frenkel and Thomas [23], as almost all others later, finally expressed their
covariantly generalized relations in terms of the usual 3-vectors
considering that the 3-vectors are physical quantities and that the UT of $%
\mathbf{E}$ and $\mathbf{B}$, $\mathbf{d}$ and\textbf{\ }$\mathbf{m}$ are
the relativistically correct LT. Particularly interesting is that Frenkel
(first and second papers in [23]) considered that $\mathbf{d}$ and\textbf{\ }%
$\mathbf{m}$ `are connected with each other by the invariant relation $%
D^{\mu \nu }u_{\nu }=0$, that is $\mathbf{d}=(1/c)\mathbf{u}\times \mathbf{m}
$,'\ (our notation) (equations (3) and (4) in the first paper in [23]),
`expressing the fact - or rather the \emph{assumption} - that in a
co-ordinate system, in which the electron's translational velocity $\mathbf{u%
}$ is zero, the `electrical moment' $\mathbf{d}$ must vanish.'\ In the
geometric approach presented here, this invariant relation would be written
as $D^{ab}u_{b}=0$, which in the standard basis becomes $D^{\mu \nu }u_{\nu
}e_{\mu }=cd^{\mu }e_{\mu }=0$. In the particle's rest frame, the $K^{\prime
}$ frame ($u^{\prime \mu }=(c,0,0,0)$), one finds that $d^{\prime 0}=0$ and $%
D^{\prime i0}=d^{\prime i}=0$, which is Frenkel's assertion, i.e. the \emph{%
assumption}. However, Frenkel's invariant relation $D^{\mu \nu }u_{\nu }=0$
is, as Frenkel says, nothing else than an \emph{assumption}, which is not
founded in any way . This means that there is no physical or mathematical
reason for the assumption that $d^{\prime i}=0$. Moreover, as already argued
several times, the 3-vectors and the relations with them, like $\mathbf{d}%
=(1/c)\mathbf{u}\times \mathbf{m}$, are not meaningful in the 4D spacetime.
Besides, in order to get from $D^{\mu \nu }u_{\nu }e_{\mu }=0$ the relation
with spatial components ($d^{i}=(1/c^{2})\varepsilon ^{0ijk}m_{j}u_{k}$),
which corresponds to Frenkel's relation between the 3-vectors $\mathbf{d}$
and\textbf{\ }$\mathbf{m}$, some additional, not justified, assumptions
(like $d^{0}=0$) are required. \bigskip \bigskip

\noindent \textbf{4. }$T$ \textbf{and} $P$ \textbf{inversions and the world
parity} $W\bigskip $

\noindent In elementary particle theories the existence of an EDM implies
the violation of the time reversal $T$ invariance. Under the assumption of
CPT invariance, a nonzero EDM would also signal CP violation. As stated in
[24], `it is the $T$ violation associated with EDMs that makes the
experimental hunt interesting.'\ Let us briefly consider the connection
between the EDM and the $T$ invariance, as it is explained in the usual
formulation, e.g. [24]. Reversing time would reverse the spin direction but
leave the EDM direction unchanged since the charge distribution does not
change. In the elementary particle theories, e.g., the standard model and
SUSY, the EDM direction is connected with a net displacement of charge along
the spin axis, i.e. with an asymmetry in the charge distribution inside a
particle; see, for example, [24]. Thus, with $t\rightarrow -t$, $\mathbf{S}%
\rightarrow -\mathbf{S}$ but $\mathbf{d}\rightarrow \mathbf{d}$. However, as
in [17], $\mathbf{d}$ is determined as $\mathbf{d}=d\mathbf{S}/S$. Hence $%
\mathbf{d}$ has to be parallel to the spin $\mathbf{S}$; it is considered
that $\mathbf{S}$ is the only available 3-vector in the rest frame of the
particle. This yields that $d\rightarrow -d$, i.e. $d\rightarrow 0$. As
stated in [24], `the alignment of spin and EDM is what leads to violations
of $T$ and $P$.'

From the viewpoint of the geometric approach presented here, neither $T$
inversion nor $P$ inversion is well-defined in the 4D spacetime; they are
not good symmetries. For the position vector $x^{a}$, only the world parity $%
W$ (for the term see, e.g., [25]), according to which $Wx^{a}=-x^{a}$, is
well-defined in the 4D spacetime. In general, the $W$ inversion cannot be
written as the product of the usual $T$ and $P$ inversions. But this will be
possible for the representations of $W$, $T$ and $P$ in the standard basis $%
\{e_{\mu }\}$. It is easy to see that, e.g., $T$ inversion is not
well-defined and that it depends, for example, on the chosen synchronization.

As explained, e.g. in [16], different systems of coordinates (including
different synchronizations) are allowed in an inertial frame and they are
all equivalent in the description of physical phenomena. Thus, in [16], two
very different but completely equivalent synchronizations, Einstein's
synchronization [5] and the `r'\ synchronization, are exposed and exploited
throughout the paper. The `r'\ synchronization is commonly used in everyday
life and not Einstein's synchronization. In the `r'\ synchronization, there
is an absolute simultaneity. As explained in [26], `For if we turn on the
radio and set our clock by the standard announcement '...`at the sound of
the last tone, it will be 12 o'clock,'\ then we have synchronized our clock
with the studio clock according to the `r'\ synchronization. In order to
treat different systems of coordinates on an equal footing we have presented
[16] the transformation matrix that connects Einstein's system of
coordinates with another system of coordinates in the same reference frame.
Furthermore, in [16] we have derived a form of the LT that is independent of
the chosen system of coordinates, including different synchronizations. The
unit vectors in the $\{e_{\mu }\}$ basis and the $\{r_{\mu }\}$ basis, i.e.
with the `r'\ synchronization , [16], are connected as
\begin{equation}
r_{0}=e_{0},\quad r_{i}=e_{0}+e_{i}.  \label{re}
\end{equation}%
Hence, the components $g_{\mu \nu ,r}$ of the metric tensor $g_{ab}$ are $%
g_{ii,r}=0$, and all other components are $=1$. Remember that in the $%
\{e_{\mu }\}$ basis $g_{\mu \nu }=diag(1,-1,-1,-1)$. (Note that in [16)] and
[15] the Minkowski metric is $g_{\mu \nu }=diag(-1,1,1,1)$.) Then, according
to (4) from [16], one can use $g_{\mu \nu ,r}$ to find the transformation
matrix $R_{\;\nu }^{\mu }$ that connects the components from the $\{e_{\mu
}\}$ basis with the components from the $\{r_{\mu }\}$ basis; $R_{\;\mu
}^{\mu }=-R_{\;i}^{0}=1$, and all other elements of $R_{\;\nu }^{\mu }$ are $%
=0$. The inverse matrix $(R_{\;\nu }^{\mu })^{-1}$ connects the `old' basis,
$\{e_{\mu }\}$, with the `new' one, $\{r_{\mu }\}$. With such an $R_{\;\nu
}^{\mu }$ one finds that the components of $x^{a}$ are connected as
\begin{equation}
x_{r}^{0}=x^{0}-x^{1}-x^{2}-x^{3},\quad x_{r}^{i}=x^{i}.  \label{ptr}
\end{equation}%
Observe that $x^{a}=x^{\mu }e_{\mu }=x_{r}^{\mu }r_{\mu }$. (Obviously, the
components of any vector transform in the same way as in (\ref{ptr}), e.g.
for the components of $E^{a}$ it holds $E_{r}^{0}=E^{0\text{ }%
}-E^{1}-E^{2}-E^{3}$, $E_{r}^{i}=E^{i}$.)\ It is clear from (\ref{ptr}) that
$T$ inversion, $t\rightarrow -t$, i.e. $x^{0}\rightarrow -x^{0}$, does not
give that $x_{r}^{0}\rightarrow -x_{r}^{0}$. This can be shown explicitly.

In the standard basis $\{e_{\mu }\}$ the matrix elements $T_{\;\nu }^{\mu }$
of the time reversal operator $T$ are $T_{\;0}^{0}=-1$, $T_{\;i}^{i}=1$ and
all other elements of $T_{\;\nu }^{\mu }$ are equal to $0$. Then, one can
write $x_{T}^{\mu }=T_{\;\nu }^{\mu }x^{\nu }$, where $x_{T}^{\mu }$ are the
components of the time reversed position vector $x_{T}^{a}=x_{T}^{\mu
}e_{\mu }$, which are $x_{T}^{0}=-x^{0}$, $x_{T}^{i}=x^{i}$. In the $%
\{r_{\mu }\}$ basis, the matrix elements $T_{\;\nu ,r}^{\mu }$ of the time
reversal operator $T$ which are different from zero are
\begin{equation}
T_{\;0,r}^{0}=-1,\quad T_{\;i,r}^{i}=1,\quad T_{\;i,r}^{0}=-2.  \label{rt}
\end{equation}%
\emph{Clearly, this is not a time reversal operation in the usual sense.} In
the $\{r_{\mu }\}$ basis, the components $x_{T,r}^{\mu }$ of the `time
reversed'\ position vector $x_{T}^{a}=x_{T,r}^{\mu }r_{\mu }$ are
\begin{equation}
x_{T,r}^{0}=-x^{0}-x^{1}-x^{2}-x^{3},\quad x_{T,r}^{i}=x^{i}  \label{tx}
\end{equation}%
and it holds that $x_{T}^{a}=x_{T}^{\mu }e_{\mu }=x_{T,r}^{\mu }r_{\mu }$.
(Of course, $x_{T,r}^{\mu }=R_{\;\nu }^{\mu }x_{T}^{\nu }=T_{\;\nu ,r}^{\mu
}x_{r}^{\nu }$.) This means that the $T$ \emph{inversion has not a definite
physical significance, since it depends on the chosen synchronization. Only
when Einstein's synchronization is used the time reversal has the usual
meaning. However, different synchronizations are nothing else than different
conventions and physics must not depend on conventions.}

In general, the same holds for the $P$ inversion. In the $\{e_{\mu }\}$
basis $P_{\;0}^{0}=1$, $P_{\;i}^{i}=-1$ and all other elements of $P_{\;\nu
}^{\mu }$ are equal to $0$. However, in the $\{r_{\mu }\}$ basis the matrix
elements $P_{\;\nu ,r}^{\mu }$ of the parity operator $P$ which are
different from zero are
\begin{equation}
P_{\;0,r}^{0}=1,\quad P_{\;i,r}^{i}=-1,\quad P_{\;i,r}^{0}=2.  \label{pe}
\end{equation}%
Obviously, \emph{in the} $\{r_{\mu }\}$ \emph{basis,} $P_{\;\nu ,r}^{\mu }$
\emph{is not a spatial inversion.} In that basis the components $%
x_{P,r}^{\mu }$ of the `spatially reversed'\ position vector $%
x_{P}^{a}=x_{P,r}^{\mu }r_{\mu }$ are
\begin{equation}
x_{P,r}^{0}=x^{0}+x^{1}+x^{2}+x^{3},\quad x_{P,r}^{i}=-x^{i}  \label{prx}
\end{equation}%
and it holds that $x_{P}^{a}=x_{P}^{\mu }e_{\mu }=x_{P,r}^{\mu }r_{\mu }$.
Thus, \emph{the parity operator} $P$ \emph{also} \emph{depends on the chosen
synchronization and therefore it is not a properly defined operation in the
4D spacetime. }$P$ \emph{has its usual meaning only when the standard basis }%
$\{e_{\mu }\}$ \emph{is chosen in some inertial frame of reference. }

On the other hand, the $W$ inversion is properly defined because if $%
x^{a}\rightarrow -x^{a}$ then $x^{\mu }\rightarrow -x^{\mu }$, $x_{r}^{\mu
}\rightarrow -x_{r}^{\mu }$, ... . Thus
\begin{eqnarray}
Wx^{a} &=&-Ix^{a};\ W_{\;\nu }^{\mu }x^{\nu }e_{\mu }=W_{\;\nu ,r}^{\mu
}x_{r}^{\nu }r_{\mu }=...=W_{\;\nu }^{\prime \mu }x^{\prime \nu }e_{\mu
}^{\prime }=W_{\;\nu ,r}^{\prime \mu }x_{r}^{\prime \nu }r_{\mu }^{\prime
}=..  \notag \\
&=&-I_{\;\nu }^{\mu }x^{\nu }e_{\mu }=-I_{\;\nu }^{\mu }x_{r}^{\nu }r_{\mu
}=...=-I_{\;\nu }^{\mu }x^{\prime \nu }e_{\mu }^{\prime }=-I_{\;\nu }^{\mu
}x_{r}^{\prime \nu }r_{\mu }^{\prime }=...,  \label{aw}
\end{eqnarray}%
where $W_{\;\nu }^{\mu }$, $W_{\;\nu ,r}^{\mu }$, $W_{\;\nu }^{\prime \mu }$%
, $W_{\;\nu ,r}^{\prime \mu }$ are the matrix elements of the proper parity
operator $W$ in the bases $\{e_{\mu }\}$, $\{r_{\mu }\}$, $\{e_{\mu
}^{\prime }\}$, $\{r_{\mu }^{\prime }\}$ and all primed quantities in (\ref%
{aw}) are Lorentz transforms of the unprimed ones; see equation (1) in [16]
for the general form of the LT. The LT in the $\{r_{\mu }\}$ basis are given
in the same paper by equation (2). The elements that are different from zero
are
\begin{eqnarray}
x_{r}^{\prime \mu } &=&L^{\mu }{}_{\nu ,r}x_{r}^{\nu },\quad
L^{0}{}_{0,r}=K,\quad L^{0}{}_{2,r}=L^{0}{}_{3,r}=K-1,\quad
L^{1}{}_{0,r}=L^{1}{}_{2,r}=  \notag \\
L^{1}{}_{3,r} &=&(-\beta _{r}/K),\quad L^{1}{}_{1,r}=1/K,\quad
L^{2}{}_{2,r}=L^{3}{}_{3,r}=1,  \label{elr}
\end{eqnarray}%
where $K=(1+2\beta _{r})^{1/2},$ and $\beta _{r}=dx_{r}^{1}/dx_{r}^{0}$ is
the velocity of the frame $S^{\prime }$ as measured by the frame $S$, $\beta
_{r}=\beta _{e}/(1-\beta _{e})$ and it ranges as $-1/2\prec \beta _{r}\prec
\infty .$ $I$ in (\ref{aw}) is the identity transformation. It can be easily
checked that $W_{\;\nu }^{\mu }=T_{\;\lambda }^{\mu }P_{\;\nu }^{\lambda
}=T_{\;\lambda ,r}^{\mu }P_{\;\nu ,r}^{\lambda }=...=-I_{\;\nu }^{\mu }$.
But the matrix elements $T_{\;\nu ,r}^{\mu }$ and $P_{\;\nu ,r}^{\mu }$,
which are given by (\ref{rt}) and (\ref{pe}), respectively, are quite
different from the usual ones from the $\{e_{\mu }\}$ basis, i.e. different
from the matrix elements of the usual $T$ and $P$ inversions. They do not
describe the time and space inversions and the notations for all, except $TP$
in the $\{e_{\mu }\}$ basis, are not adequate.

It is worth noting that the same relations as in (\ref{aw}) hold also for $%
d^{a}$, $Z^{a}$, $m^{a}$, $S^{a}$, $E^{a}$, $B^{a}$, etc., i.e., $%
Wd^{a}=-Id^{a}$, ... . \emph{One law for the proper inversion }$W$ \emph{for
all vectors!} Hence, $L_{int}$ from (\ref{i}) is unchanged under the proper
inversion $W$. The $W$ inversion is well-defined symmetry in the 4D
spacetime. This is drastically different from the usual $T$ and $P$
inversions for the 3-vectors. For the $T$ inversion of $\mathbf{d}$ and $%
\mathbf{S}$, see, e.g., the beginning of this section.

This fact that $T$ and $P$ inversions are not well-defined symmetries in the
4D spacetime is one of the reasons why, contrary to the existing elementary
particle theories, the $T$ violation, i.e. the $CP$ violation, cannot be
connected in this approach with the existence of an intrinsic EDM.

Another reason is that, as already stated, neither the direction of $\mathbf{%
d}$ nor the direction of the spin $\mathbf{S}$ have a well-defined meaning
in the 4D spacetime. The only Lorentz-invariant condition on the directions
of $d^{a}$ and $S^{a}$ in the 4D spacetime is $d^{a}u_{a}=S^{a}u_{a}=0$.
This condition does not say that $\mathbf{d}$ has to be parallel to the spin
$\mathbf{S}$. The above discussion additionally proves that the relations (%
\ref{es}), (\ref{dm}) and (\ref{gu}) are properly defined.

If an antisymmetric tensor (the components) $A^{\mu \nu }$ (that tensor $%
A^{ab}$ can be, e.g., $F^{ab}$, $M^{ab}$, $S^{ab}$, $D^{ab}$, ...) is
transformed by $R_{\;\nu }^{\mu }$ to the $\{r_{\mu }\}$ basis, then it is
obtained that
\begin{equation}
A_{r}^{10}=A^{10}-A^{12}-A^{13},  \label{are}
\end{equation}%
which shows that the `time-space'\ components in the $\{r_{\mu }\}$ basis
are expressed by the mixture of the `time-space'\ components and the
`space-space'\ components from the $\{e_{\mu }\}$ basis. Thus, for example,
\begin{equation}
D_{r}^{10}=-d^{1}+(1/c)m^{3}-(1/c)m^{2}.  \label{dre}
\end{equation}%
Similarly, it follows from (\ref{are}) that $F_{r}^{10}=E^{1}+cB^{3}-cB^{2}$%
. The relation (\ref{dre}) and the one for $F_{r}^{10}$ show, once again,
that \emph{the components have no definite physical meaning}. Only in the $%
\{e_{\mu }\}$, $\{e_{\mu }^{\prime }\}$ bases does it hold that $%
E^{i}=F^{i0} $, $d^{\prime i}=D^{\prime i0}$, $Z^{\prime i}=S^{\prime i0}$,
etc. That is the reason why we always put the quotation marks in the
expressions `time-space'\ and `space-space.'\ One important consequence of (%
\ref{are}) and (\ref{dre}) is that the usual EDMs and MDMs, $\mathbf{d}$ and
$\mathbf{m} $, respectively, where, e.g., $\mathbf{d=}D^{10}\mathbf{i}+D^{20}%
\mathbf{j}+D^{30}\mathbf{k}$, \emph{have no definite physical meaning, }%
since the components $D^{i0}$ are dependent on the chosen synchronization.
Of course, the same holds for the fields $\mathbf{E}$ and $\mathbf{B}$. In
contrast with the usual covariant approach with coordinate-dependent
quantities, all relations (\ref{f})-(\ref{dm}) are written in terms of 4D
geometric quantities, i.e. they are defined without reference frames. This
means that dipole moments $d^{a}$ and $m^{a}$ are well-defined quantities in
the 4D spacetime but, according to (\ref{d}), they depend not only on $%
D^{ab} $ but also on the 4-velocity of the particle $u^{a}$ as well. Hence,
as already stated, $D^{ab}$ is the primary quantity; it does not depend on $%
u^{a}$. The same assertion can be stated for the relation between $F^{ab}$
and $E^{a}$, $B^{a}$, $v^{a}$, as seen from (\ref{f}).

All this proves that in the `r'\ synchronization it is not possible to speak
about time and space as separate quantities. So, in the 4D spacetime, $W$
inversion has \textit{an independent reality }in Minkowski's sense but%
\textit{\ }not $T$ and $P$ inversions. Similarly, $D^{ab}$ has \textit{an
independent reality }but\textit{\ }not the dipole moments $\mathbf{d}$ and $%
\mathbf{m}$. The same applies to $F^{ab}$ and the fields $\mathbf{E}$ and $%
\mathbf{B}$. By the explicit use of the `r'\ synchronization I have
mathematically formalized Minkowski's words. Note that only in Einstein's
synchronization are the spatial and temporal parts of the interval between
the two spacetime points separated. The usual covariant approaches
implicitly use only Einstein's synchronization and therefore the majority of
physicists believe that $T$ and $P$ inversions taken separately are
well-defined symmetries. A similar conclusion applies to $\mathbf{d}$ and $%
\mathbf{m}$ and the fields $\mathbf{E}$ and $\mathbf{B}$.\bigskip \bigskip

\noindent \textbf{5. Shortcomings in the EDM searches\bigskip }

\noindent The results obtained offer some new interpretation of measurements
of an EDM of a fundamental particle, e.g. [8, 24, 27]. In all experimental
searches for a permanent EDM of particles, the UT of $\mathbf{E}$ and $%
\mathbf{B}$ are frequently used and considered to be relativistically
correct; that is, that they are the LT of $\mathbf{E}$ and $\mathbf{B}$.
Thus in a recent new method of measuring EDMs in storage rings [8], the
so-called motional electric field, our $\mathbf{E}^{\prime }$, is considered
to arise `according to a Lorentz transformation'\ from a vertical magnetic
field $\mathbf{B}$ that exists in the laboratory frame; $\mathbf{E}^{\prime
}=\gamma c\mathbf{\beta \times B}$. That field $\mathbf{E}^{\prime }$ plays
a decisive role in the mentioned new method of measuring EDMs. It is stated
in [8] that $\mathbf{E}^{\prime }$ `can be much larger than any practical
applied electric field.'\ and `Its action on the particle supplies the
radial centripetal force.'\ Then, after introducing `g-2'\ frequency $\omega
_{a}$ ($\omega _{a}=a(eB/m)$, $a=(g-2)/2$ is the magnetic anomaly) due to
the action of the magnetic field on the muon magnetic moment, they say, `If
there is an EDM of magnitude $d=\eta $%
h{\hskip-.2em}\llap{\protect\rule[1.1ex]{.325em}{.1ex}}{\hskip.2em}%
$e/4mc\simeq \eta \times 4.7\times 10^{-14}ecm$, there will be an additional
precession angular frequency $\mathbf{\omega }_{e}=(\eta e/2m)\mathbf{\beta
\times B}$ about the direction of $\mathbf{E}^{\prime }$, ... .'\ The new
technique of measuring EDM in [8] is to cancel $\omega _{a}$ so that $%
\mathbf{\omega }_{e}$ can operate by itself. An important remark on such a
treatment is that the field $\mathbf{E}^{\prime }$ is in the rest frame of
the particle $K^{\prime }$ but the measurement of EDM is in the laboratory
frame $K$. A similar thing happens in [27] and many others in which
`motional magnetic field' $\mathbf{B}^{\prime }=(\gamma /c)\mathbf{E\times
\beta }$ appears in the particle's rest frame as a result of the UT of the $%
\mathbf{E}$ field from the laboratory. It is usually considered that the $%
(\gamma /c)\mathbf{E\times \beta }$ field causes important systematic
errors. Thus, it is stated already in the abstract in the first paper in
[27]: `In order to achieve the target sensitivities it will be necessary to
deal with the systematic error resulting from the interaction of the
well-known $\mathbf{v\times E}$ field with magnetic field gradients .. .
This interaction produces a frequency shift linear in the electric field,
mimicking an EDM.'\ The same interpretation with the UT of $\mathbf{E}$ and $%
\mathbf{B}$ appears when the quantum phase of a moving dipole is considered,
e.g. [28]. For example, when the R\"{o}ntgen phase shift is considered, it
is asserted in the second paper in [28] that in `the particle rest frame the
magnetic flux density $\mathbf{B}$ due to the magnetic line is perceived as
an electric field'\ $\mathbf{E}^{\prime }=\mathbf{v}\times \mathbf{B}$. Then
that $\mathbf{E}^{\prime }$ can interact with $\mathbf{d}^{\prime }$ in $%
K^{\prime }$. This is objected to in [7]. In the usual approaches with the
3-vectors it is also possible to get the interaction between $\mathbf{B}$
and $\mathbf{d}$ by another method, which conforms more to a description in $%
K$. According to the second method, the magnetic field $\mathbf{B}$ in $K$
interacts with the MDM $\mathbf{m}$ that is obtained from the EDM $\mathbf{d}%
^{\prime }$ by the UT for $\mathbf{m}$ and $\mathbf{d}$; $\mathbf{m}=\gamma
\mathbf{v\times d}^{\prime }$. For the Aharonov-Casher effect, this method
is mentioned in, e.g. [29]. However, as already said, the transformations of
$\mathbf{E}$ and $\mathbf{B}$, (\ref{bx}), and of $\mathbf{d}$ and $\mathbf{m%
}$, are not the LT but the UT [11]. They have to be replaced by the LT of
the corresponding 4D geometric quantities. Then, the LT transform $B^{\mu }$
from $K$ again to $B^{\prime \mu }$ in $K^{\prime }$ and, similarly, $E^{\mu
}$ from $K$ is transformed again to $E^{\prime \mu }$ in $K^{\prime }$ (\ref%
{be}); there is no mixing of components. The same holds for the LT of $%
d^{\mu }$ and $m^{\mu }$. Thus, in this approach, there is no induced $%
\mathbf{E}^{\prime }$ as in [8] and [28], and there is no `motional magnetic
field' $\mathbf{B}^{\prime }$ as in [27] and [29], and there is no induced $%
\mathbf{d}$ in $K$ as in [29].

As already mentioned, in all EDM experiments the interaction between the
electromagnetic field and the dipole moments is described in terms of the
3-vectors as $\mathbf{E}\cdot \mathbf{d}$ and $\mathbf{B}\cdot \mathbf{m}$.
Moreover, the 3-vectors $\mathbf{d}$ and $\mathbf{m}$ (and also $\mathbf{E}$
and $\mathbf{B}$) are in the rest frame of the particle, whereas the
measurements of EDM are in the laboratory frame. However, the last two terms
in (\ref{i}) show that in the geometric approach presented here there are
direct interactions between the magnetic field $B^{a}$ and an EDM $d^{a}$
and also between $E^{a}$ and $m^{a}$, which are required for the explanation
of measurements in [8] and [27]. In order to describe the interactions in $K$
one only needs to choose the laboratory frame $K$ as the $e_{0}$-frame and
then to represent $E^{a}$, $B^{a}$ and $d^{a}$, $m^{a}$ in that frame. This
can be explained in more detail comparing the expression (\ref{i}) with the
interaction term in the Lagrangian in, e.g., equation (17) in [17], which is%
\begin{equation}
L_{int}=\mathbf{d}^{\prime }\cdot \mathbf{E}^{\prime }\mathbf{+B}^{\prime
}\cdot \mathbf{m}^{\prime }+(1/c^{2})\mathbf{u\cdot m}^{\prime }\mathbf{%
\times E}^{\prime }-\mathbf{u\cdot d}^{\prime }\mathbf{\times B}^{\prime }.
\label{an}
\end{equation}%
This is written in our notation in which the particle's rest frame is the $%
K^{\prime }$ frame and the quantities from that frame are the primed
quantities. (There is an ambiguity with the notation. Namely, $L_{int}$ (\ref%
{an}) is in $K^{\prime }$, but at the same time that expression contains the
particle's velocity $\mathbf{u}$.)

Since the measurements are in the laboratory frame (the $K$ frame) we shall
first choose the laboratory frame as the $e_{0}$-frame and represent $E^{a}$%
, $m^{a}$, $B^{a}$, $d^{a}$, ... from (\ref{i}) in that frame. Hence, in the
laboratory frame, which is taken to be the $e_{0}$-frame, $v^{\mu
}=(c,0,0,0) $ and consequently $E^{0}=B^{0}=0$; see (\ref{f}). Then (\ref{i}%
) becomes
\begin{eqnarray}
L_{int} &=&-((E_{i}d^{i})+(B_{i}m^{i}))-(1/c^{2})\varepsilon
^{0ijk}(E_{i}m_{k}-c^{2}B_{i}d_{k})u_{j}  \notag \\
&&+(1/c)((E_{i}u^{i})d^{0}+(B_{i}u^{i})m^{0}).  \label{lg}
\end{eqnarray}%
Observe that in the laboratory frame there are contributions from the terms
with $d^{0}$ and $m^{0}$. The contribution of the terms with the interaction
of $E^{a}$ with $m^{a}$ ($B^{a}$ with $d^{a}$) is $u/c$ of the usual terms
with the direct interaction of $E^{a}$ with $d^{a}$ ($B^{a}$ with $m^{a}$).
The constraints $d^{a}u_{a}=m^{a}u_{a}=0$ also can be written in the $e_{0}$%
-frame, which enables us to express $d^{0}$ and $m^{0}$ by means of $%
d^{i}u_{i}$ and $m^{i}u_{i}$, respectively. Then, it can be seen that terms
with $d^{0}$ and $m^{0}$ in (\ref{lg}) are $u^{2}/c^{2}$ of the usual terms $%
E_{i}d^{i}$ or $B_{i}m^{i}$ and therefore they can be neglected.

In the usual approach with the 3-vectors (neglecting terms of the order of $%
u^{2}/c^{2}$) $L_{int}$ (\ref{lg}) would correspond to
\begin{equation}
L_{int}=\mathbf{d}\cdot \mathbf{E+m}\cdot \mathbf{B}+(1/c^{2})\mathbf{u\cdot
m\times E}-\mathbf{u\cdot d\times B},  \label{d3}
\end{equation}%
where, in contrast with (\ref{an}), all quantities are in the laboratory
frame. It is assumed that the spatial components with upper indices from (%
\ref{lg}) correspond to the spatial components of the 3-vectors in (\ref{d3}%
). Namely, the metric is diag$(1,-1,-1,-1)$ and $\varepsilon ^{0123}=1$.

In an unrealistic case when the rest frame of the dipole is chosen to be the
$e_{0}$-frame, i.e. when the observers who measure fields $E^{a}$ and $B^{a}$
move together with the dipole, $v^{a}=u^{a}=ce_{0}^{\prime }$ and
consequently $E^{\prime 0}=B^{\prime 0}=d^{\prime 0}=m^{\prime 0}=0$, then $%
L_{int}$ from (\ref{i}) becomes
\begin{equation}
L_{int}=-(E_{i}^{\prime }d^{\prime i})-(B_{i}^{\prime }m^{\prime i}).
\label{lc}
\end{equation}%
In the usual picture with 3-vectors it would correspond to
\begin{equation}
L_{int}=\mathbf{d}^{\prime }\cdot \mathbf{E}^{\prime }\mathbf{+m}^{\prime
}\cdot \mathbf{B}^{\prime }.  \label{l3d}
\end{equation}%
The corresponding Hamiltonian is
\begin{equation}
H_{int}=-\mathbf{d}^{\prime }\cdot \mathbf{E}^{\prime }\mathbf{-m}^{\prime
}\cdot \mathbf{B}^{\prime }.  \label{hi}
\end{equation}%
$H_{int}$ (\ref{hi}) is the form of the Hamiltonian of the interaction that
is frequently used in the comparison with the standard theory and in the
interpretation of the results of EDM experiments. $\mathbf{E}^{\prime }$%
\textbf{\ }and\textbf{\ }$\mathbf{B}^{\prime }$ in (\ref{hi}) are both in
the rest frame of the dipole and in all the usual approaches they are
expressed in terms of the laboratory fields using the UT of $\mathbf{E}$%
\textbf{\ }and\textbf{\ }$\mathbf{B}$, (\ref{bx}). Observe, once again, that
$L_{int}$ (\ref{l3d}) and $H_{int}$ (\ref{hi}) refer to the case when the
observer who measures fields $\mathbf{E}^{\prime }$\textbf{\ }and\textbf{\ }$%
\mathbf{B}^{\prime }$ `sits'\ on the moving dipole.

In the 4D geometric approach presented in this paper, expressions like (\ref%
{d3}), (\ref{l3d}) and (\ref{hi}) are meaningless because, as explained
particularly in [12], there are not the usual time-dependent 3-vectors in
the 4D spacetime. The relativistically correct 4D expressions are (\ref{lg})
in the laboratory frame (or, neglecting terms of the order of $u^{2}/c^{2}$,
$L_{int}$ (\ref{lg}) without last two terms) and (\ref{lc}) in the rest
frame of the dipole when that frame is at the same time the $e_{0}$-frame.
They are derived from (\ref{i}), while the Lagrangian $L_{int}$ (\ref{i}) is
obtained using mathematically and physically correct decompositions (\ref{f}%
) and (\ref{d}).

For the phase shifts these questions are discussed in [3] and [7].
Accordingly, the experimentalists who search for an EDM, e.g. [8] and [27],
and, for example, those who observe the Aharonov-Casher phase shift [30],
will need to reexamine the results of their measurements taking into account
the relations (\ref{i}) and (\ref{es})- (\ref{dm}). \bigskip \bigskip

\noindent \textbf{6. Conclusion}

\noindent In conclusion, we believe that the new results (\ref{es})-(\ref{dm}%
) that are obtained in this paper, together with the expression (\ref{i})
for the interaction term, [3], provide an alternative but viable formulation
of spins and dipole moments. It will be of interest in different branches of
physics, particularly elementary particle theories and experiments, and also
theories and experiments that treat different quantum phase shifts with
dipoles. It is worth noting that the relations (\ref{em}), (\ref{es}) and (%
\ref{dm}) are generalized to the quantum case and the new commutation
relations for the orbital and intrinsic angular momentums and for the dipole
moments are introduced in [31]. \bigskip \bigskip

\noindent \textbf{References\bigskip }

\noindent \lbrack 1] Ivezi\'{c} T 2007 \textit{Found. Phys.} \textbf{37} 747

\noindent \lbrack 2] Ivezi\'{c} T 2006 \textit{Found. Phys.} \textbf{36} 1511

Ivezi\'{c} T 2007 \textit{Fizika A (Zagreb)} \textbf{16} 4

\noindent \lbrack 3] Ivezi\'{c} T 2007 \textit{Phys. Rev. Lett.} \textbf{98}
108901

\noindent \lbrack 4] Ivezi\'{c} T 2005 \textit{Found. Phys. Lett.} \textbf{18%
} 401

\noindent \lbrack 5] Einstein A 1905 \textit{Ann. Physik.} \textbf{17} 891
translated by Perrett W and Jeffery

G B 1952 in \textit{The Principle of Relativity} (New York: Dover)

\noindent \lbrack 6] Ludvigsen M 1999 \textit{General Relativity,} \textit{A
Geometric Approach }

(Cambridge: Cambridge University Press)

Sonego S and Abramowicz M A 1998 \textit{J. Math. Phys.} \textbf{39} 3158

Hillion P 1993 \textit{Phys. Rev. E} \textbf{48} 3060

\noindent \lbrack 7] Ivezi\'{c} T 2007 \textit{Phys. Rev. Lett.} \textbf{98}
158901

\noindent \lbrack 8] Farley F J M, Jungmann K, Miller J P, Morse W M, Orlov
Y F,

Roberts B L, Semertzidis Y K, Silenko A and Stephenson E J

2004 \textit{Phys. Rev. Lett.} \textbf{93} 052001

\noindent \lbrack 9] Jackson J D 2004 \textit{Am. J. Phys.} \textbf{72} 1484

\noindent \lbrack 10] Jackson J D 1998 \textit{Classical Electrodynamics}
3rd edn (New York: Wiley)

\noindent \lbrack 11] Ivezi\'{c} T 2003 \textit{Found. Phys.} \textbf{33}
1339\textbf{\ }

Ivezi\'{c} T 2005 \textit{Found. Phys.} Lett. \textbf{18 }301\textbf{\ }

Ivezi\'{c} T 2005 \textit{Found. Phys.} \textbf{35} 1585

Ivezi\'{c} T 2008 \textit{Fizika A }\textbf{17} 1

\noindent \lbrack 12] Oziewicz Z 2008 \textit{Rev. Bull. of the Calcutta
Math. Soc. }\textbf{16} 49

Oziewicz Z and Whitney C K 2008 \textit{Proc. Natural Philosophy Alliance }

\textit{(NPA)} \textbf{5} 183 (also at http://www.worldnpa.org/php/)

Oziewicz Z 2009 Unpublished results that can be obtained

from the author at oziewicz@unam.mx

\noindent \lbrack 13] Ivezi\'{c} T 2009 arXiv: 0906.3166

\noindent \lbrack 14] Minkowski H 1908 \textit{Nachr. Ges. Wiss. G\"{o}%
ttingen} 53

(The French translation by Paul Langevin is at

http://hal.archives-ouvertes.fr/hal-00321285/fr/

the Russian translation is in 1983 \textit{Einshteinovskii Sbornik} 1978-1979

(Moskva: Nauka) pp. 5-63

\noindent \lbrack 15] Ivezi\'{c} T 2002 \textit{Found. Phys. Lett.} \textbf{%
15} 27

Ivezi\'{c} T 2001 arXiv: \textit{physics}/0103026

Ivezi\'{c} T 2001 arXiv: \textit{physics}/0101091

\noindent \lbrack 16] Ivezi\'{c} T 2001 \textit{Found. Phys.} \textbf{31}
1139

\noindent \lbrack 17] Anandan J 2000 \textit{Phys. Rev. Lett.} \textbf{85}
1354

\noindent \lbrack 18] Uhlenbeck G E and Goudsmit S 1925 \textit{Naturwissen}
\textbf{13 }953

Uhlenbeck G E and Goudsmit S 1926 \textit{Nature} \textbf{117} 264

\noindent \lbrack 19] Westpfahl K 1967 \textit{Ann. Physik.} \textbf{20} 113

\noindent \lbrack 20] Westpfahl K 1967 \textit{Ann. Physik.} \textbf{20} 241

\noindent \lbrack 21] Wang Z Y and Xiong C D 2006 \textit{Chin. Phys.}
\textbf{15} 2223

\noindent \lbrack 22] Peletminskii A and Peletminskii S 2005 \textit{Eur.
Phys. J. C} \textbf{42} 505

\noindent \lbrack 23] Frenkel J 1926 \textit{Nature }\textbf{117} 653

Frenkel J 1926 \textit{Z. Physik} \textbf{37} 243

Thomas L H 1926 \textit{Nature }\textbf{117} 514

Thomas L H 1927 \textit{Phil. Mag.} \textbf{3} 1

\noindent \lbrack 24] Fortson N, Sandars P and Barr S 2003 \textit{Physics
Today} \textbf{56} 33

\noindent \lbrack 25] Arunasalam V 1994 \textit{Found. Phys. Lett.} \textbf{7%
} 515

Arunasalam V 1997 \textit{Phys. Essays} \textbf{10} 528

\noindent \lbrack 26] Leubner C, Aufinger K and Krumm P 1992 \textit{Eur. J.
Phys.} \textbf{13} 170

\noindent \lbrack 27] Barabanov A L, Golub R and Lamoreaux S K 2006 \textit{%
Phys. Rev. }

\textit{A} \textbf{74} 052115

Baker C A et al. 2006 \textit{Phys. Rev. Lett.} \textbf{97} 131801

Regan B C, Commins E D, Schmidt C J and DeMille D 2002

\textit{Phys. Rev. Lett.} \textbf{88} 071805

\noindent \lbrack 28] Wilkens M 1994 \textit{Phys. Rev. Lett.} \textbf{72 }5

Horsley S A R and Babiker M 2005 \textit{Phys. Rev. Lett.} \textbf{95} 010405

\noindent \lbrack 29] Zeilinger A, G\"{a}hler R and Horne M A 1991 \textit{%
Phys. Lett. A} \textbf{154} 93

\noindent \lbrack 30] Cimmino A, Opat G I, Klein A G, Kaiser H, Werner S A,
Arif M

and Clothier R 1989 \textit{Phys. Rev. Lett.} \textbf{63} 380

Sangster K, Hinds E A, Barnett S, Riis E and Sinclair A G

1995 \textit{Phys. Rev. A} \textbf{51} 1776

\noindent \lbrack 31] Ivezi\'{c} T 2007 arXiv: hep-th/0705.0744

\end{document}